\documentstyle[12pt]{article}
\begin{document}
{
\begin{center}
{\Large \bf The singlet-triplet magnetism and induced spin
fluctuations \\
in the high-$T_c$ copper oxides.}

\vspace{0.5cm}
{A.S. Moskvin$^{\star}$ and A.S. Ovchinnikov}

\vspace{0.3cm}
{
\normalsize
Department of Theoretical Physics, Ural State University,\\
620083, Lenin Ave. 51, Ekaterinburg, Russia.
}
\end{center}
\vspace{0.5cm}

{\normalsize
High-$T_c$ cuprates like $La_{2-x}Sr_{x}CuO_{4}$ and $YBa_{2-x}Cu_{3}O_{6+x}$ are considered as a system of the electron and hole polar pseudo-Jahn-Teller $CuO_{4}$ centers $[{CuO}_{4}^{7-}]_{JT}$ and $[CuO_{4}^{5-}]_{JT}$, respectively, or a system of the local bosons moving in a lattice of the hole centers. Ground manifold of the polar centers includes three terms $^{1}A_{1g}$ (Zhang-Rice singlet), $^{1}E_{u}$, $^{3}E_{u}$ with different spin multiplicity, orbital degeneracy and parity that provides an unconventional multi-mode behaviour of the cuprates.
The spin subsystem  of the copper oxides within the polar Jahn-Teller $CuO_{4}$ centers model is a two-component spin liquid and corresponds to a singlet-triplet magnet with, in general, multiparametric noncollinear spin configurations. In the framework of a modified mean field approximation some kinds of spin ordering are discussed including as a trivial singlet or triplet states as a pure quantum singlet-triplet mixed state.
A local boson movement is accompanied by a modulation of the spin density on the site resulting in the so called induced spin fluctuations. Some unconventional features of the induced spin fluctuations are considered including an appearance of the induced longitudinal ferrimagnetism with an appropriate contribution to the spin susceptibility, a possibility to observe and examine the charge fluctuations with the help of the traditional magnetic methods such as the magnetic inelastic neutron scattering and the spin lattice relaxation experiments. The suggested model, in comparison with  the nearly antiferromagnetic Fermi-liquid model, represents  new approach to the description of the spin system of the high-$T_c$ cuprates.
}

\vspace{0.5cm}

PACS codes/keywords: 74.20.Hi, 74.25.Ha


\vspace{1cm}

$^{\star}$Corresponding author: Prof. A.S. Moskvin,  Department of Theoretical Physics,   Ural State University, 620083, Ekaterinburg, Russia\\
 Fax: +7-3432-615-978 	\,\,\, E-mail: alexandr.moskvin@usu.ru

\newpage{}

\section{Introduction.}

Unconventional properties of the oxides like  cuprates ($YBa_2Cu_3O_{6+x}$, $La_{2-x}Sr_xCuO_4$, $La_2CuO_{4+d}$,..\\.), manganites ($La_{1-x}Sr_xMnO_3$,...), nickellates ($La_2NiO_{4+d}$,...), bismuthates ($(K,Ba)BiO_3$) including systems with the high-$T_c$ superconductivity and colossal magnetoresistance  reflect a result of a response of the system to the nonisovalent substitution or charge disorder that stabilizes the intermediate valence phases providing the most effective screening of the charge inhomogeneity. These phases in oxides may involve novel molecular cluster configurations like the Jahn-Teller  $sp$-center   \cite{Moskvin}  with anomalously high local polarizability. Their appearance may be a result of the disproportionation reaction like
 \[
                     M +M = M^{+} + M^{-},
\]
                                                                              where $M$ is a basic metal-oxygen center ($CuO_4^{6-}$, $NiO_4^{6-}$, $MnO_6^{9-}$, $BiO_6^{8-}$, respectively), the $M^\pm$ centers are the corresponding polar (hole and electron) centers. All these oxides are the so called charge transfer semiconductors, where a fundamental absorption band is determined by the charge transfer $s\rightarrow p$ transition from the nondegenerate and  predominantly metallic even $s$-state to the low lying degenerate and  predominantly oxygen odd $p$-state. An ionization or a hole doping for the $sp$-like $M$-center may be accompanied by the quasi-degeneracy effect: a hole may be localized either at the predominantly metallic ($s$) or at the predominantly oxygen ($p$) molecular orbital with competition of two configurations, $s^2$ and $sp$, respectively. As a result, we have for the hole center $M^-$  a ground manifold of the terms with different parity, spin multiplicity and orbital degeneracy providing a multi-mode behavior of such  centers. First of all, these are unstable with respect to the pseudo-Jahn-Teller effect  \cite{Moskvin,Moskvin1,Moskvin2} with  active even and odd local displacement modes of different symmetry and formation of the appropriate vibronic states  \cite{Ber}.
The hole pseudo-Jahn-Teller (PJT) center with its high polarizability can be a center of an effective local pairing with a formation of the local singlet (e.g. cuprates) or triplet (e.g. manganites) boson as  two electrons paired in molecular shell. Thus, we come to the electron PJT center: $M^- = M^+ + (\mbox{local boson})$. A local boson can correspond to the completely filled molecular shell with the $^{1}A_{1g}$ symmetry (e.g. $b_{1g}^2$  for the cuprates and nickellites and $6s^2$  for the bismuthates) or to the half-filled molecular shell $e_g^2$ with the $^{3}A_{2g}$ symmetry in manganites. Two polar centers may have a similar structure of the ground manifold. It is worth to note that the PJT nature of the polar centers results in a strong vibronic reduction for a probability of their recombination.
  As a whole the oxides under such consideration may be called as {\it the strongly correlated PJT oxides}.

  In particular, a model approach developed in papers \cite{Moskvin,Moskvin1,Moskvin2} considers the $CuO_{4}$ cluster based copper oxides as systems unstable with respect to the disproportionation reaction
\[
	2 {CuO}_{4}^{6-} \to  [{CuO}_{4}^{5-}]_{PJT} + [CuO_{4}^{7-}]_{PJT}
\]
with the creation of the system of the polar hole
($h$) $CuO_{4}^{5-}$ or electron ($e$) $CuO_{4}^{7-}$ pseudo-Jahn-Teller  centers. These centers are distinguished by the local boson or by two electrons paired in the completely filled molecular orbital of the $CuO_{4}$-cluster. The new phase can be considered as a system of the local bosons moving in the lattice of the $h$-centers or as a generalized quantum lattice bose-gas. An origin and anomalous properties of the $h$-centers are connected with a near degeneracy of the molecular terms $^{1}A_{1g}$ (Zhang-Rice singlet), $^{1}E_{u}$, $^{3}E_{u}$ for the configurations $b_{1g}^{2}$ and $b_{1g}e_{u}$, respectively, that can create conditions for the pseudo-Jahn-Teller  effect with active local displacements modes of the $Q_{eu}$, $Q_{b_{1g}}$ and $Q_{b_{2g}}$ types.
    In general, the PJT effect in the $^{1}A_{1g}$, $^{1,3}E_{u}$ manifold leads to the formation of the four-well adiabatic potential of two symmetry types: $B_{1g}$ or $B_{2g}$ with the nonzero local displacements of the $Q_{eu}$, $Q_{b_{1g}}$ or $Q_{eu}$, $Q_{b_{2g}}$-types, respectively (Fig.1). Unusual properties of the $^{1}A_{1g}$, $^{1,3}E_{u}$ manifold with the terms distinguished by the spin multiplicity, parity and orbital degeneracy provide unconventional behaviour for the $h$-centers with an active interplay of various modes.  The hole PJT center with its high polarizability can be a center of an effective local pairing.

A most direct and convincing observation of the hole singlet-triplet PJT centers is made recently  \cite{Fisk} by the NQR method in $La_{2}Cu_{0.5}Li_{0.5}O_{4}$.
	 The authors revealed the spin singlet ground state ($S=0$) and the low lying spin triplet state ($S=1$) with the singlet-triplet separation $\Delta_{ST} = 0.13 eV$ comparable with the well known estimates for the nearest neighbours exchange interactions for the parent cuprates. They found the anomalously weak temperature dependence of the relaxation rate at low temperatures that evidences an occurrence of the spinless multiplet structure in the  $CuO_{4}$ cluster ground state. They revealed an appreciable spin contribution to the low temperature relaxation indicating the simultaneous occurrence of the ground state multiplet structure, the sufficiently low singlet-triplet separation and the intrinsic singlet-triplet mixing. These features and the observed  relaxation inequivalence of the various Cu sites are quite natural for the PJT centers in the conditions of the static JT-effect.

The PJT hole centers like $CuO_{4}^{5-}$ with the singlet-triplet quasi-degeneracy within a ground state have been observed by ESR-spectroscopy in $LaSrAl_{1-x}Cu_{x}O_4$ which is isostructural to $La_{2-x}Sr_xCuO_4$ \cite{Kazan}.

A description of the multimode PJT centers system is very complicated. The simplest model  Hamiltonians correspond to the quantum lattice bose-gas Hamiltonian  for the charge subsystem  \cite{robasz,Alexandrov1} and the Hamiltonian of the cooperative JT-problem for the vibronic subsystem  \cite{Ber}.  Below we'll present a simplest model approach to the spin subsystem of the PJT centers phase.

Unconventional properties of the spin subsystem in the PJT centers phase as for the generalized singlet-triplet magnetic system with well developed charge, spin and structure fluctuations are connected with {\it the induced  spin fluctuations}. The main source of them is determined by a dependence of the singlet-triplet separation $\Delta_{ST}$ on the local boson density.
 The local boson movement is accompanied by the fluctuations of the $\Delta_{ST}$ value and appropriate spin fluctuations. In general, for a spectral region of the spin system transparency (or far from the main spin excitations) the temporal dependence of the induced spin fluctuations  may be entirely determined by the  charge fluctuations.
This  enables to explain many unusual spectral and momentum $(\omega, \vec q)$, temperature and concentration $(x,T)$ dependences for different effects determined by the spin correlation functions including an inelastic neutron scattering (INS) and the spin-lattice relaxation.
Moreover, it indicates a principal possibility to make use of the magnetic methods (INS, NMR, spin susceptibility measurements) for the revealing and studying the non-magnetic (charge and vibronic)  excitations.

The paper is organized as follows. In Section II we consider the mean field approximation (MFA) for the simplified singlet-triplet Hamiltonian of the PJT centers system with total neglect the intermode coupling. In Section III we suggest a model approach to account of the spin-charge correlations within a so called induced spin fluctuations model. Some static and dynamical spin properties for various copper oxides are discussed in Section IV in a framework of the singlet-triplet  model.

\section{The singlet-triplet model of the polar PJT centers phase. Mean field analysis.}

To maximally simplify the spin subsystem description within the PJT centers phase we'll restrict ourselves only to the singlet and triplet ${}^{1}E_{u}$, ${}^{3}E_{u}$  terms for the $b_{1g}e_{u}$-configuration (see Fig.2) neglecting the orbital degeneracy  and taking into account  the isotropic spin exchange interaction between  $b_{1g}$ and $e_{u}$ holes of the neighbouring $CuO_{4}$ centers. Thus, we come to the Hamiltonian of the two-component spin liquid
\[
  H_{ST}= \frac{1}{2} \sum_{i} I_{b_{1g}e_{u}}(ii){\vec s}_{1i}{\vec s}_{2i}
    +\frac{1}{2} \sum_{ij} I_{b_{1g}b_{1g}}(ij){\vec s}_{1i}{\vec s}_{1j}+
     \frac{1}{2} \sum_{ij} I_{b_{1g}e_{u}}(ij){\vec s}_{1i}{\vec s}_{2j}+
\]
\begin{equation}
   +  \frac{1}{2} \sum_{ij} I_{e_{u} b_{1g}}(ij){\vec s}_{2i}{\vec s}_{1j}+
     \frac{1}{2} \sum_{ij} I_{e_{u} e_{u}}(ij){\vec s}_{2i}{\vec s}_{2j},
\end{equation}
where the ${\vec s}_{1i}$, ${\vec s}_{2i}$ are the spins localized on the predominantly copper $b_{1g}$ or the pure oxygen $e_{u}$ orbitals, respectively. First term in (1) describes the intra-center $dp$-exchange and the rest ones describe the inter-center $dd,\,dp,\,pp$-exchange, respectively. Introducing the summary spin of $i$-th cluster
\[
{\vec S}_{i} = {\vec s}_{1i}+{\vec s}_{2i}
\]
 and the spin polarization operator
\[
{\vec V}_{i} = {\vec s}_{1i}-{\vec s}_{2i}
\]
 with the non-zero matrix element like $\langle 00 | V_{z}| 10 \rangle = 1$ responsible for the singlet-triplet mixing one can rewrite (1) as
\[
  H_{ST}=
	\frac{\triangle _{ST}}{2} \sum_{i} {\vec S}_{i}^{2}+
	\frac{1}{2} \sum_{ij} I_{ij}^{(1)}{\vec S}_{i}{\vec S}_{j}+
     \frac{1}{2} \sum_{ij} I_{ij}^{(2)}{\vec S}_{i}{\vec V}_{j}+
\]
\begin{equation}
  +  \frac{1}{2} \sum_{ij} I_{ij}^{(3)}{\vec V}_{i}{\vec S}_{j}+
     \frac{1}{2} \sum_{ij} I_{ij}^{(4)}{\vec V}_{i}{\vec V}_{j}.
\end{equation}
Here we used the new exchange integrals
\begin{equation}
  I_{ij}^{(1)}=\frac{1}{4}\lbrace I_{b_{1g}b_{1g}}(ij) + I_{e_{u}e_{u}}(ij) + 2I_{b_{1g}e_{u}}(ij) \rbrace,
\end{equation}
\begin{equation}
  I_{ij}^{(2)}=I_{ij}^{(3)}=\frac{1}{4}\lbrace I_{b_{1g}b_{1g}}(ij) - I_{e_{u}e_{u}}(ij) \rbrace,
\end{equation}
\begin{equation}
  I_{ij}^{(4)}=\frac{1}{4}\lbrace I_{b_{1g}b_{1g}}(ij) + I_{e_{u}e_{u}}(ij) - 2I_{b_{1g}e_{u}}(ij) \rbrace
\end{equation}
and also introduced  the singlet-triplet separation $\triangle _{ST}=I_{b_{1g}e_{u}}(ii)$. The first term in (2) explicitly takes account of the intra-center $dp$-exchange resulting in the singlet-triplet separation for each cluster, the second one describes the conventional  Heisenberg exchange of triplets leading to the formation of spin waves (SW). The rest terms are non-trivial and correspond to the  important   effects of  spin polarization exchange with possible formation of magnetic excitons (spin multiplicity waves) and  their hybridization with SW.

Numerical values and signs of the exchange parameters $I^{(1,2,3,4)}$ are determined as a result of a competition among the $b_{1g}-b_{1g}$, $b_{1g}-e_{u}$ and $e_{u}-e_{u}$ exchange interactions. Take notice of different signs  of the $b_{1g}-e_{u}$ contribution to
$I^{(1)}$ and $I^{(4)}$ and also  of the $b_{1g}-b_{1g}$ and $e_{u}-e_{u}$ contributions to $I^{(2)}=I^{(3)}$.

Contrary to the well known weakly covalent and weakly correlated magnetic  oxides like ferrites, the exchange interaction for two nearest neighboring PJT centers in the strongly covalent cuprates \cite{Bourges}  includes the unusually large ferromagnetic Hund (for different orbitals) or antiferromagnetic correlational (for similar orbitals) contributions of the intra-atomic $O_{2p}$-exchange within the common oxygen ion. Besides, we should deal with a relatively large ferromagnetic contribution of the Heisenberg $Cu3d-O2p$ exchange. As a whole, a detailed microscopic analysis of the exchange interactions for the PJT centers requires a special consideration, especially, in conditions of their comparable large values $I\sim \Delta _{ST} \sim 0.1\,eV$.

Notice that some kinds of the singlet-triplet Hamiltonians have been discussed to describe the magnetic behaviour of some rare-earth magnetic systems with $Pr^{3+}$, $Tb^{3+}$, $Tm^{3+}$ ions  \cite{birgeneau,Hsieh}. Recently, the singlet-triplet magnet model has been suggested for $La_{2}NiO_{4}$ compound  \cite{oles}.

Above we neglect the nonmagnetic Zhang-Rice singlet ${}^{1}A_{1g}$  which inclusion in the $H_{ST}$ leads to an essential complication with a renormalization of the exchange parameters. Note only an appearance of the supplementary contribution like
\[
\sum_{ij} U_{ij} {\vec S}^{2}_{i} {\vec S}^{2}_{j}
\]
 as a result of the inter-center electrostatic interactions $U$ with account  of the different electron density distribution for singlet ${}^{1}A_{1g}$ and triplet ${}^{3}E_{u}$ states.  This term leads to the possibility of the singlet-triplet ordering (like the charge one) without any spin-exchange interactions.

The Hamiltonian (2) has the $SO(4)$ group symmetry  \cite{Bohm}
\begin{equation}
   [S_{i},S_{j}] = i {\varepsilon}_{ijk} S_{k}, \qquad
   [S_{i},V_{j}] = i {\varepsilon}_{ijk} V_{k}, \qquad
   [V_{i},V_{j}] = i {\varepsilon}_{ijk} S_{k},
\end{equation}
an algebra of which has two invariant operators (Casimir's operators)
\begin{equation}
   C^{SO(4)}_{1} = {\vec S}^{2} + {\vec V}^{2} = 3 {\bar I}, \qquad
   C^{SO(4)}_{2} = {\vec S} \cdot {\vec V} =0.
\end{equation}
These relations provide a kinematic constraint for $S$ and $V$ modes and automatically follow from the ordinary spin algebra.

Below, we'll treat the Hamiltonian (2) in the framework of the modified  MFA
approach within the nearest neighbours approximation with two sublattices labelled by  additional $\gamma = A, B$ indices. Then Hamiltonian (2) can be rewritten as
\begin{equation}
   H_{MFA}=\frac{\triangle}{2} \sum_{i \gamma} {\vec S}_{i \gamma}^{2}-\frac{1}{2}\sum_{i \alpha \gamma} h_{s i \alpha \gamma} \langle S_{i \alpha}
\rangle_{\gamma}-\frac{1}{2} \sum_{i \alpha \gamma} h_{v i \alpha \gamma} \langle V_{i \alpha} \rangle_{\gamma}+
\end{equation}
\begin{equation}
+ \sum_{i \alpha \gamma} h_{s i \alpha \gamma} S_{i \alpha \gamma}+
\sum_{i \alpha \gamma} h_{v i \alpha \gamma} V_{i \alpha \gamma}
\end{equation}
with molecular fields
\begin{equation}
   h_{s i \alpha \gamma} = \sum_{j, {\gamma}^{'}} (I_{ij}^{(1)}{\langle S_{j {\alpha} }\rangle}_{{\gamma}^{'}}+ I_{ij}^{(2)}{\langle V_{j {\alpha}}\rangle}_{{\gamma}^{'}}),
\end{equation}
\begin{equation}
  h_{v i \alpha \gamma} = \sum_{j, {\gamma}^{'}} (I_{ij}^{(2)} {\langle S_{j{\alpha}}\rangle}_{{\gamma}^{'}} + I_{ij}^{(4)}
{\langle V_{j {\alpha}}\rangle}_{{\gamma}^{'}}).
\end{equation}

Note, that the MFA treatment of the Hamiltonians (1) and (2) is different in what concerns the taking account of the first term describing the intra-center exchange interaction. Contrary to the usual MFA,  a summary spin $SM_S$-representation used in  the Hamiltonian (2) provides an explicit account of the intra-center exchange. Within the usual MFA approach for the Hamiltonian (1) the averages $\langle \vec S \rangle$ and $\langle \vec V \rangle$ (often denoted as $\vec M$ and $\vec L$) correspond to the intra-center "ferromagnetic" and "antiferromagnetic" vectors for the PJT center, respectively, with kinematic constraint
\[
{\langle {\vec S} \rangle}^{2} + {\langle {\vec V} \rangle}^{2} = 4s_{1}^{2}=4s_{2}^{2},\qquad {\langle {\vec S} \rangle} \cdot {\langle {\vec V} \rangle} =0
\]
instead of the operator constraint (7) which in the appropriate averages is equivalent to
\[
{\langle {\vec S}^{2} \rangle} + {\langle {\vec V}^{2} \rangle} = 4 s_{1}(s_{1}+1)=3,\qquad {\langle {\vec S} \cdot {\vec V} \rangle} =0.
\]
So, the singlet-triplet model within the modified MFA approach permits more correct account of the quantum effects essential at low spin magnitudes $s_{1,2}=\frac{1}{2}$.

An occurrence of two vector order parameters even with kinematic constraint (7) makes an analysis extremely complicated.
The relatively simple collinear spin configurations like those that depicted in Fig.3(b,c), could be obtained only for four particular cases: 1) $I^{(1,2,3,4)}<0$, $\Theta _{AA}=\Theta _{BB}=\Theta _{AB}=0$, 2) $I^{(1,4)}>0$, $I^{(2,3)}<0$, $\Theta _{AA}=\Theta _{BB}=\Theta _{AB}=\pi $, 3) $I^{(1,2,3,4)}>0$, $\Theta _{AA}=\Theta _{BB}=0$, $\Theta _{AB}=\pi$, 4) $I^{(1,4)}<0$, $I^{(2,3)}>0$, $\Theta _{AA}=\Theta _{BB}=\pi$, $\Theta _{AB}=0$, as direct analogues of the ferro- and antiferromagnetic ordering.
Take notice of the corresponding $SS$- and $VV$-configurations could be obtained as the usual MFA solutions of the Hamiltonian (1) only at $T=0$ when the MFA constraint ${\langle {\vec S} \rangle} \cdot {\langle {\vec V} \rangle} =0$ is fulfilled both for the $SS$ (${\langle V \rangle}=0$) and the $VV$ ($\langle S \rangle = 0$) phase.

 In general, we have to deal with a frustration of the $SS,VV,SV$ type exchange interactions that implies an appearance of the two-sublattice multiparametric ($\langle S \rangle, \langle V \rangle, \Theta _{AA} = \Theta _{BB}, \Theta _{AB}$)  angular $S,V$ spin configurations schematically depicted in Fig.3(a).
Their occurrence extends substantially a set of admissible spin states and results in effective involving of the spin system to the strong inter-mode coupling and hybridization that raises its role in the general optimization procedure, especially, with taking account of the large inhomogeneity.

Leaving the detailed MFA analysis of the phase diagram for the singlet-triplet Hamiltonian (2) extremely complicated for the angular modes to the separate consideration  we'll consider at first a specific case of the collinear $S,V$ structures.

In the nearest neighbours approximation for the collinear spin structures the free energy per site is
\[
  {\tilde F}= \frac{F}{N} = -\frac{1}{4} \sum_{\gamma = A,B}(h_{sz\gamma} {\langle S_{z} \rangle}_{\gamma}+
h_{vz \gamma} {\langle V_{z} \rangle}_{\gamma}) + \frac{\Delta}{2} -
\]
\begin{equation}
- \frac{1}{2 \beta} \sum_{\gamma =A, B}\ln{\{ 2(\cosh(\frac{\beta}{2} {\alpha}_{\gamma})+\exp{(-\beta \frac{\triangle}{2})} \cosh{(\beta {h}_{sz \gamma})})\}}.
\end{equation}
Here  $\alpha_{\gamma}={\lbrace {\triangle}^2+4{h}^{2}_{vz \gamma} \rbrace}^{\frac{1}{2}}$, $\beta = \frac{1}{T}$ denotes the inverse of the temperature.

The set of self-consistent equations for the order parameters minimizing  ${\tilde F}$ is
\begin{equation}
	{\langle S_{z} \rangle}_{\gamma}=-\frac{\exp{(-\beta \frac{\triangle}{2})} \sinh{(\beta h_{s z \gamma})}}{\cosh{(\frac{\beta}{2} {\alpha}_{\gamma})}+ \exp{(-\beta \frac{\triangle}{2})} \cosh{(\beta h_{s z \gamma})}},
\end{equation}
\begin{equation}
	{\langle V_{z} \rangle}_{\gamma}=-\frac{2 \frac{{h}_{vz\gamma}}{{\alpha}_{\gamma}} \sinh{(\frac{\beta}{2} {\alpha}_{\gamma})}}{\cosh{(\frac{\beta}{2} {\alpha}_{\gamma})}+ \exp{(-\beta \frac{\triangle}{2})} \cosh{(\beta h_{s z \gamma})}} .
\end{equation}

For $T=0$ this system gets a form
\[
{{\langle V_{z} \rangle}_{\gamma}}=
\Biggl \{
\begin{array}{ccc}
-2 \frac{{h}_{vz\gamma}}{{\alpha}_{\gamma}} & , & \frac{1}{2}{\alpha}_{\gamma} > |{h}_{s z \gamma}|- \frac{\triangle}{2}	\\
0  &, &	\frac{1}{2}{\alpha}_{\gamma} <  |{h}_{s z \gamma}|- \frac{\triangle}{2} \\
\end{array},
\]

\[
{{\langle S_{z}\rangle}_{\gamma}}=
\Biggl \{
\begin{array}{ccc}
-\mbox{sgn}{{h}_{s z \gamma}} & , & \frac{1}{2}{\alpha}_{\gamma} < |{h}_{s z \gamma}|- \frac{\triangle}{2}	\\
0  &, &	\frac{1}{2}{\alpha}_{\gamma} > |{h}_{s z \gamma}|- \frac{\triangle}{2} \\
\end{array}.
\]

Obviously, these equations have the following non-trivial solutions:

1) {\bf SS-phase}: (${\langle S_{z} \rangle}_{A}={\eta}_{1} \not= 0 $,
${\langle S_{z} \rangle}_{B}={\eta}_{2} \not= 0 $,
${\langle V_{z} \rangle}_{A} = {\langle V_{z} \rangle}_{B} = 0)$, where $\eta_{1}=-\eta_{2}=\pm 1$ ($I^{(1)}>0$) or $\eta_{1}=\eta_{2}=\pm 1$ ($I^{(1)}<0$).  This solution is realized when
\begin{equation}
   \triangle < 2|h_{s z \gamma}|-{\alpha}_{\gamma}  \qquad (\gamma = A, B)
\end{equation}
or with account of the appropriate order parameter values
\begin{equation}
  \triangle + {\lbrace {\triangle}^2+{(2 z I^{(2)})}^2 \rbrace}^{\frac{1}{2}} < 2 z |I^{(1)}|.
\end{equation}
(Here and below z is the nearest neighbours number, $I^{(1,2,4)}$
is an effective nearest neighbours exchange integral).
This relation means   that one of the pure triplet states $| 1-1 \rangle$ or $| 11 \rangle $ has the minimal energy  at both sublattices $A, B$.

2) {\bf VV-phase}: (${\langle V_{z} \rangle}_{A}={\eta}_{1} \not= 0 $, ${\langle V_{z} \rangle}_{B}={\eta}_{2} \not= 0 $,  ${\langle S_{z} \rangle}_{A} = {\langle S_{z} \rangle}_{B} = 0)$. By similar way we obtain
$
   \eta_{1} = \eta_{2}=\eta \qquad (I^{(4)}<0); \qquad
	 \eta_{1} = - \eta_{2}=\eta \qquad (I^{(4)}>0);
$
\begin{equation}
   |\eta|={\lbrace 1- (\frac{\triangle}{2zI^{(4)}})^2\rbrace}^{\frac{1}{2}}.
\end{equation}

This kind of solutions is realized if
\begin{equation}
   \triangle > 2|h _{s z \gamma}|-{\alpha}_{\gamma} \qquad  (\gamma = A, B),
\end{equation}
that corresponds to nonequality
\begin{equation}
	|I^{(4)}|{\lbrace 2z|I^{(4)}|+\triangle \rbrace}>|I^{(2)}|{\lbrace (2zI^{(4)})^{2}-{\triangle}^{2} \rbrace}^{\frac{1}{2}}.
\end{equation}
It means that on each site $A$ or $B$ the state minimizing the energy is a quantum mixture of the $| 00 \rangle$ and $| 10 \rangle$ states.
It is worth to note that the situation widely explored in theoretical investigations  (well isolated singlet) is the particular version of the $VV$-phase. Only when the singlet-triplet separation is large enough compared with an effective magnetic polarisation exchange ($\Delta > 2 z I^{(4)} $) the spin subsystem of the new phase of the polar PJT centers has the singlet as a ground state.

Note again, that within the MFA approach the $\vec S$ and $\vec V$ vectors correspond to the "ferromagnetic" and "antiferromagnetic" vectors for the $CuO_{4}$-center, respectively. Fig.4 shows qualitatively an orientation of the copper and oxygen spins for different $S,V$ arrangements.

Above we have considered the various $SS$ and $VV$ modes at zero temperature. At $T \not= 0$ we have, in general, the nonzero values both for $\langle S_{z} \rangle$ and for $\langle V_{z} \rangle$ or, in other words, the phase mixing ($SS-VV$) occurres. The unconventional non-Brillouin temperature behaviour of the appropriate order parameters and  spin susceptibilities for the obtained phases are presented in Fig.5,6 together with the simplified spin level schemes for the antiferromagnetic $SS$ and $VV$ ordering (exchange parameters  $I^{(1,4)}$, $\triangle _{ST}=I_{b_{1g}e_{u}}(ii)$ are chosen to be positive, exchange parameters  $I^{(2,3)}$ are chosen to be negative).

  The Curie (Neel) temperature $T_{c}$ is defined as a temperature when all spin order parameters tend to  zero simultaneously. Expanding the system (13-14) on the corresponding small order parameters we get the pair of nonlinear equations
\[
	2 \exp{(\beta \frac{\triangle}{2})} \sinh{(\beta \frac{\triangle}{2})}
{\lbrace (2zI^{(2)})^{2} - 4 z^2 I^{(1)} I^{(4)} \rbrace}=(3+ \exp{(\beta \triangle)}) \times
\]
\begin{equation}
{\lbrace
  \frac{\triangle}{\beta} (3+ \exp{(\beta \triangle)}) \pm 2 z I^{(1)}{\triangle} \pm 4 \frac{z I^{(4)}}{\beta}{\exp{(\beta \frac{\triangle}{2})}} \sinh{(\beta \frac{\triangle}{2})}
\rbrace},
\end{equation}
from solution of which we get $T_{1,2}$, then $T_{c} = \max \{ T_{1}, T_{2} \}$.

An existence of two spin order parameters leads to many unconventional properties of the cuprates. First of all, outline  various possible types of magnetic arrangement in the system of the PJT polar centers. Along with the well studied phase of  (anti)ferromagnetic ordering of the Neel type ($\langle S_{z} \rangle \not= 0$, $\langle V_{z} \rangle = 0$) but with the non-Brillouin temperature dependence the new phases can occur. Only when $\langle S_{z} \rangle = \langle V_{z} \rangle = 0$, $\langle S^{2} \rangle = 0$, $\langle V^{2} \rangle = 3$ ($T=0$)  all polar centers have the singlet ground state.
 Non-trivial phases are the solutions with the non-zero order parameter $\langle V_{z} \rangle$ when $\langle S_{z} \rangle =0$.
These correspond to different spin density arrangements.

Unusual low temperature dependences of the order parameters and susceptibilities may be revealed by the singlet-triplet system in the $SS-VV$ crossover regime, when
\[
   \triangle \approx 2|h _{s z \gamma}|-{\alpha}_{\gamma} \qquad  (\gamma = A, B).
\]

An occurrence of two spin order parameters results in two types of the spin susceptibilities: the usual one
\[
\chi _{S}\sim \frac{\partial \langle \vec S \rangle}{\partial \vec h}
\]
and a new spin-polarisation susceptibility
\[
\chi _{V}\sim \frac{\partial \langle \vec V \rangle}{\partial \vec h}
\]
with a principally different temperature behavior presented in Figs 5,6. Note that  the MFA predicts new spin-polarization susceptibility $\chi _{V}$ should turn into zero within paramagnetic region as a consequence of the zero's values of the corresponding molecular fields and an absence of the singlet-triplet mixing. So, experimental reveal of the nonzero $\chi _{V}$ will indicate a relative role of the $VV$-correlations.

Let note that the taking account of the singlet-triplet mixing or the $V$-terms results in a suppression of the "$S$-type" magnetism with a peculiar quantum "nearly paramagnetic" behavior of the system. This reflects both in a behavior of the $S$-order parameter and in a tendency to coincidence of the transversal and longitudinal susceptibilities.

The partial copper and oxygen spin polarizations within the PJT center and the appropriate susceptibilities can have unconventional behavior for the considered phases determined by the antiferromagnetic $SS$ and $VV$ exchange interactions. So, with increasing the temperature for the $SS$-phase  we have a crossover from the ferromagnetic intra-center ordering ($\langle V_{z} \rangle \sim 0$) to the state with intensive quantum suppression of the copper polarization ($\langle V_{z} \rangle \sim - \langle S_{z} \rangle$). In the $VV$-phase a similar suppression occurs beginning from the antiferromagnetic intra-center ordering. Both phases reveal unconventional behavior of the copper $\chi _{d}=\frac{1}{2}(\chi _{S}+\chi _{V})$ and oxygen $\chi _{p}=\frac{1}{2}(\chi _{S}-\chi _{V})$ susceptibilities. The close values and the same signs of the  $\chi _{S}$ and $\chi _{V}$ result in a  large magnitude of the former  and a near cancellation of the latter. Naturally, in paramagnetic state both susceptibilities coincide.

\section{Noncollinear phase.}

In considered model the principal feature of noncollinear spin configurations  is  their two-sublattice nature with combined ferro-aniferromagnet arrangement of vector order parameter components $\langle \vec S \rangle$ and $\langle \vec V \rangle$ (see Fig.3(a)) and wave vector $\vec Q_{AF}=(\pi,\pi)$. This fact distinguishes them from the usual incommensurate noncollinear spin structures.

The analytical consideration of angular phase is cumbersome and hardly  advisable therefore we discuss only the numerical results. The temperature behaviour of longitudinal and transversal order parameters for noncollinear  phase is presented in Fig.7(a)  (for the sake of convinience we have choosed $z$ axis along ${\langle S_{z} \rangle}_{A}$ vector).  In contrast to collinear phases, in considered  situation both order parameters  $\langle \vec S \rangle$ and $\langle \vec V \rangle$ are nonzero simultaneously at $T=0$ though the orthogonality condition
$$
		\langle {\vec S} \cdot {\vec V} \rangle = 0
$$
is satisfied for each site.

The further temperature evolution of spin arrangement resembles the well known orientational transition in antiferromagnets. In Fig.7(b)  we present
the temperature dependence of $\Theta_{AA}=\Theta_{BB}$ angles between
$\langle \vec S \rangle$ and $\langle \vec V \rangle$ vectors on $A,B$ sites, $\Theta_{AB}$ between ${\langle \vec S \rangle}_{A}$ and ${\langle \vec S \rangle}_{B}$ vectors. From this plot one can see that with increasing the temperature the sublattice turning occurs when the spin parameters attempt   to  be collinear
${\langle \vec S \rangle}_{A} \parallel {\langle \vec S \rangle}_{B}$ and ${\langle \vec V \rangle}_{A} \parallel {\langle \vec V \rangle}_{B}$
(explicit type of arrangement is determined by the signs of exchange interactions  $I^{(1,2,4)}$).
Beginning from the temperature   $T^{\star}$ on-site spin order parameters $\langle \vec S \rangle$ and $\langle \vec V \rangle$ line up along $z$ axis that is accompanied by sharp decreasing to zero of appropriate transversal components. It should be noticed that for noncollinear phase  a clear difference between longitudinal $\chi_{s}^{\parallel}$ (in Fig.8 applied magnetic field $\vec H \parallel z$) and transversal  $\chi_{s}^{\perp}$ (in Fig.8 $\vec H \perp z$) susceptibilities disappears that determines nearly paramagnetic behaviour of system up to low temperatures and looks like spin-glass behaviour.

As for collinear phases $\chi_{v}^{\parallel}$, $\chi_{v}^{\perp}$  turn into zero at Neel temperature. We pay attention to anomalous temperature behaviour of the spin polarization susceptibilities $\chi_{v}^{\parallel,\perp}$. At natural positive $\chi_{s}^{\parallel,\perp}$
the spin polarization susceptibility $\chi_{v}^{\parallel}$ changes sign at $T \sim 0.18 eV$ and $\chi_{v}^{\perp}$ is negative everywhere.

In connexion of full energy optimization, an appearance of multiparametric  angular spin configurations instead of the rigid collinear (anti)ferromagnetic structures essentially extends the possibilities of spin subsystem in cuprates , especially, in conditions of strong inhomogeneity   of intra- and intercenter exchange parameters and complicates an observation of magnetic arrangement in experiments.

\section{Induced spin fluctuations.}

Of course, the above considered singlet-triplet Hamiltonian within the mean field  approximation gives an oversimplified  picture of the real situation in cuprates and should  be applied  to the real systems with some caution. The developed singlet-triplet model corresponds actually to the {\it weak inter-mode coupling regime} for the cuprates with relatively independent charge, spin and vibronic subsystems. Within this model approach an inter-mode coupling can be reduced to the renormalization of the parameters such as the transfer integral, the exchange integrals etc. So, in full analogy with vibronic reduction \cite{isotope} we could introduce a {\it spin reduction} of the local spinless boson  transfer integral resulting in more or less effective suppression of the critical temperature $T_c$. Note that within a {\it weak inter-mode coupling regime} this suppression is determined only by the dependence of the spin state of the PJT center on the boson density and is slightly dependent on the character of spin-spin correlations. In general, a magnitude of the {\it spin reduction} of the local  boson  transfer integral and its dependence on different factors such as concentration can essentially influence the superconducting properties of the cuprates.

In general, for the cuprates we should account for the strong intermode spin-charge-vibronic coupling   and an occurence of the strong charge and structure fluctuations resulting in a suppression of the  long range spin order. Actually, we deal with complicated multi-component strongly correlated quantum bose-liquid with the hybrid charge-spin-local structure fluctuations providing the most effective screening of the charge inhomogeneity.
Nevertheless, the cuprates can exhibit the various short range spin fluctuations revealed in the magnetic  inelastic neutron scattering and the spin relaxation phenomena.

Among the various intermode coupling effects we mention shortly an influence  of the vibronic PJT-modes on the exchange parameters and in more details a dependence of the singlet-triplet separation $\Delta _{ST}$  on the boson density.

Fig.2 clearly indicates the strong dependence of the exchange parameters on the vibronic modes that can be expressed within a pseudospin formalism, for  example, as
\[
I(mn) = I + I_{1}({\hat \sigma}_{z}(m)+
{\hat \sigma}_{z}(n)) +
I_{2} {\hat \sigma}_{z}(m) {\hat \sigma}_{z}(n) +
I_{orb}{\hat l_{z}}(m) {\hat l_{z}}(n)
\]
where ${\hat {\vec l}}(m)$ is the Izing like orbital moment occuring for the $e_u$-hole. Parameters $I_{1,2}$ and $I_{orb}$ are determined by the $e_{u}$-hole contribution.  Thus, the spin fluctuations should be strongly coupled not only to the vibronic or local structure quadrupole fluctuations but to pair local orbital current fluctuations.

Quantitative account of the strong intermode coupling in cuprates is an extremely complicated challenging problem which needs in some model simplifications.
 Below, within the so called induced spin fluctuations approximation we'll consider in details the role of  the well developed static and dynamic charge fluctuations.
  Within the local spinless bosons model and {\it weak inter-mode coupling regime} their effect on the spin subsystem is mainly determined by the dependence of the magnitude of the singlet-triplet separation $\Delta_{ST}$ on the boson density distribution
\begin{equation}
	\Delta_{ST}(m)={\bar \Delta}_{ST}(m)+\sum_{n} {\Delta}(mn){\delta N}(n),
\end{equation}
where ${\bar \Delta}_{ST}(m)$ is the magnitude of the singlet-triplet separation for the $m$ site in the absence of the charge fluctuations, $\delta N (n) = N(n) - \langle N (n) \rangle$ is a boson density (number) fluctuation. The second term in (21) with ${\Delta}(mn)$ is a contribution to the singlet-triplet separation for the $m$ site determined by the charge fluctuations on the $n$ site.

Parameters $\Delta (mn)$ at $m = n$ describe the local intra-center correlation effects and those at $m\not= n$ describe the nonlocal crystalline field effects. Within the above mentioned simplified two-terms ($^{1,3}E_u$) model the singlet-triplet separation
\[
\Delta _{ST}= E(^{3}E_u ) -  E(^{1}E_u )
\]
has to be slightly dependent on the crystalline field effect and an appropriate non-local contribution to $\Delta _{ST}$ is rather small. Within a real three-terms ($^{1}A_{1g}, ^{1,3}E_u$) model a ground singlet is a vibronic mixture of the singlet terms $^{1}A_{1g}$ and $^{1}E_u$, so the singlet-triplet separation $\Delta_{ST}$ should be strongly dependent on the crystalline field effects. The fluctuational contribution in the crystalline field parameters can be represented in terms of   the $partial$ crystalline field parameters
\begin{equation}
	\Delta B_{kq}= \sum_{n} b_{kq}({\vec R}_{mn}){\delta N}(n),
\end{equation}
that, in their turn, can be reliably evaluated, for instance, within a point charge model
\begin{equation}
	b_{kq}(\vec R_{mn})=\frac{2e^2}{R^{k+1}_{mn}}C^{k}_{q}(\vec R_{mn}).
\end{equation}

The fluctuations of the singlet-triplet separation result in the appropriate spin fluctuations for which in the framework of the mean field approximation we'll obtain a relatively simple linear coupling
\begin{equation}
	\delta {\vec S}_{m}(t)=\sum_{n}(p^{s}_{mn}{\vec h}^{s}_{m}(t) + p^{v}_{mn}{\vec h}^{v}_{m}(t)) \delta {\vec N}_{n}(t),
\end{equation}
where ${\vec h}^{s, v}$ is a molecular field of the S or V type, respectively.
The known expressions for the molecular fields (10)-(11) enables us to couple the spin fluctuations $\delta {\vec S}_{m}(t)$ with the $\vec S$ and $\vec V$ operators
\begin{equation}
	\delta {\vec S}_{m}(t)=\sum_{n}(P^{s}_{mn}{\vec S}_{m}(t) + P^{v}_{mn}{\vec V}_{m}(t))
\delta {\vec N}_{n}(t),
\end{equation}
where the quantities $P^{s,v}_{mn}$ represent actually the spin-charge susceptibilities. Molecular fields and spins in the right hand side of the expressions (24) and (25) have been taken at the average boson density without an account of charge fluctuations.

The spin-charge susceptibilities
 $p^{s,v}_{mn}$ and $P^{s,v}_{mn}$, in general, can  vary within sufficiently wide range. In this connection, let outline the specific case of the so called optimized systems with identical structure of the low lying terms for  the electron and hole PJT centers \cite{isotope}. Within this "rigid ground multiplet" approximation $\Delta (mn)=0$, $P^{s,v}_{mn}=0$ and $\delta {\vec S}_{m}(t)=0$.

The spin fluctuations $\delta {\vec S}_{m}(t)$ determined by the charge fluctuations in accordance with (25)  will be further called as {\bf the induced spin fluctuations}. These fluctuations reflect the charge-spin mode hybridization that determines their unconventional $\vec q$-dependence even for the simple charge-spin modes with the spin structure like spin density wave (SDW)
\begin{equation}
	{\vec S}_{m}(t)={\vec S}_{0}(t) \cos{({\vec q}_{sp}{\vec R}_{m})},
\end{equation}
specified by the wave vector $ \pm {\vec q}_{sp}$, and, in general, for the incommensurate charge structure like charge density wave (CDW)
\begin{equation}
	{\delta N}_{m}(t)={\delta N}_{0}(t) \cos{({\vec q}_{ch}{\vec R}_{m}+\alpha)},
\end{equation}
where $\vec q_{ch}$ is a specific wave vector and  $\alpha$ is a relative CDW-SDW phase shift.

In common case, we can convert (25) to the expression
\begin{equation}
	\delta {\vec S}_{\vec q}(t)= \frac{1}{N} \sum_{\vec Q}
P^{s}_{\vec Q}{\vec S}_{\vec q - \vec Q}(t)
\delta N_{\vec Q}(t)
\end{equation}
with certain relation for the appropriate $\vec q$ components.
(Here and further we restrict ourservels with the "$S$-type" contribution. The corresponding account of the "$V$-type" contribution is trivial.)

Below, we assume this relation is valid not only for classical quantities  $\vec S$,
$\delta N$, $\delta {\vec S}$ but for the corresponding operators.
The appropriate correlation functions for the induced spin fluctuations can be represented  as
\begin{equation}
	\langle \delta {\hat {\vec S}}_{{\vec q}}(0) \delta {\hat {\vec S}}_{-{\vec q}}(t) \rangle =
	\frac{1}{N^{2}}
	\sum_{{\vec Q},{\vec Q}'} P^{s}_{\vec Q}P^{s}_{-{\vec Q}'} \times
	{\langle
	{\hat {\vec S}}_{{\vec q} - {\vec Q}}(0)
	{\hat {\vec S}}_{- {\vec q} + {\vec Q}^{'}}(t)
        {\delta \hat N}_{\vec Q}(0)
        {\delta \hat N}_{-{\vec Q}^{'}}(t)
	\rangle} ,
\end{equation}
where the hybrid spin-charge correlation functions are presented.
An approximation of weakly interacting charge and spin subsystems for  (29)  implies a breaking apart  the hybrid correlation functions, i.e.
\begin{equation}
\langle {\hat {\vec S}}_{{\vec q} - {\vec Q}}(0) {\hat {\vec S}}_{-{\vec q}+ {\vec Q}'}(t)
 {\delta {\hat N}}_{{\vec Q}}(0) {\delta {\hat N}}_{-{\vec Q}'}(t) \rangle \approx
\langle {\hat {\vec S}}_{{\vec q} - {\vec Q}}(0) {\hat {\vec S}}_{-{\vec q}+ {\vec Q}'}(t)\rangle
 \langle {\delta {\hat N}}_{{\vec Q}}(0) {\delta {\hat N}}_{-{\vec Q}'}(t) \rangle .
\end{equation}

An appearance of the hybrid correlation functions in (29) directly points to the possibility to detect the non-magnetic  charge dynamic fluctuations by means of the magnetic methods (inelastic neutron scattering, spin resonance, nuclear resonance, etc.). Let note an appearance of the critical behavior of the charge fluctuations near the charge ordering temperature $T_{CO}$ within the local bosons system (see Fig.9). This unconventional behavior of the charge system will be revealed in  unusual temperature behavior of the  induced spin fluctuations and in the appropriate spin properties. In particular, note the pseudo-gap behavior of  the  induced spin fluctuations below the temperature $T_{CO}$ of the charge ordering in the boson system.

Above we restricted ourselves with a case of the single $CuO_2$-plane cuprates like
$La_{2-x}Sr_{x}CuO_{4}$ with one $CuO_{4}$-center per unit cell. For the "bi-plane" cuprates like  $YBa_{2}Cu_{3}O_{6+x}$ it is neccessary to take account of two $CuO_{4}$-centers per unit cell. Then modified expression (29) for the induced spin fluctuations takes the form
$$
\langle \delta {\hat {\vec S^{\mu }}}_{{\vec q}}(0)
\delta {\hat {\vec S^{\mu ^{'}}}}_{-{\vec q}}(t) \rangle =
\frac{1}{N^{2}} \sum _{{\vec Q}, {\vec Q}^{'}}
\sum _{\nu, {\nu}^{'}=1,2}
e^{i {\vec Q}{\vec \rho}_{\mu \nu}}
e^{-i {\vec Q}'{\vec \rho}_{\mu ' \nu '}}
P^{s}(\vec Q) P^{s}(-{\vec Q}') \times
$$
\begin{equation}
\langle
{\hat {\vec S^{\mu}}}_{\vec q - \vec Q}(0) {\hat {\vec S^{\mu'}}}_{-\vec q + {\vec Q}^{'}}(t)
{\delta {\hat N}}^{\nu}_{\vec Q}(0) {\delta {\hat N}}^{\nu^{'}}_{-{\vec Q}'}(t)
\rangle  ,
\end{equation}

where indices ${\mu}$, ${\mu}^{'}$ (${\nu}$, ${\nu}^{'}$) numerate the sublattices; ${\vec \rho}_{\mu \nu} = \vec r_{\mu} - \vec r_{\nu}$ ($\vec r_{i}$ is a vector indicating the $CuO_4$ centers within unit cell):
$$
P^{s}(\vec q) = \sum_{{\vec \delta} = \vec R_{mn}+{\vec \rho}_{\mu \nu}}
P^{s}_{m \mu; n \nu} e^{-i {\vec q} {\vec \delta}}
$$
is the Fourier component of the spin-charge susceptibility.

Above expressions (24) - (31) enable to uncover a series of the unconventional properties of the  induced spin fluctuations. Firstly, we note that the dynamics of the  induced spin fluctuations can not only reflect but be determined by the dynamics of the charge fluctuations.
This effect displays itself distinctly in the spin subsystem transparency region far from the main spin excitations, when we can neglect the temporal dependence of the spin correlation function in the expressions (29)-(31). Then, the temporal dependence of  the  induced spin fluctuations will be entirely determined by that of the charge fluctuations.

Let's draw attention to the unconventional contribution in the  induced spin fluctuations connected with the special  for the singlet-triplet magnetic system one-center averages like
 $\langle {\hat {\vec S}}^{2}_{m} \rangle$ ($\langle {\hat {\vec V}}^{2}_{m} \rangle$) or in the $q$-representation
\begin{equation}
	\langle {\hat {\vec S}}_{\vec q} {\hat {\vec S}}_{-{\vec q}'}\rangle = \delta_{\vec q {\vec q}'} \langle {\hat {\vec S}}^{2} \rangle  .
\end{equation}

In accordance with the common expression (29) at  $\vec Q = \vec Q'$,
\begin{equation}
\langle \delta {\hat {\vec S}}_{\vec q}(0) \delta {\hat {\vec S}}_{-{\vec q}}(t) \rangle =
\frac{1}{N}
 \sum_{\vec Q} |P^{s}_{\vec Q}|^{2} \langle {\hat {\vec S}^{2}} \rangle
\langle {\delta {\hat N}}_{\vec Q}(0) {\delta {\hat N}}_{-{\vec Q}}(t) \rangle ,
\end{equation}
so  the appropriate term in the   induced spin fluctuations is independent on $q$ and its $t$-dependence will be entirely determined by that of the charge fluctuations.

The  induced spin fluctuations result in unconventional contribution $\Delta \chi$ to the static magnetic susceptibility determined by the correlation function
\begin{equation}
	\langle \delta {\hat {\vec S}}_{0}(0) \delta {\hat {\vec S}}_{0}(0) \rangle \approx
\frac{1}{N}
\sum_{\vec Q} |P^{S}(\vec Q)|^{2} \times
\langle {\hat {\vec S}}_{\vec Q}(0) {\hat {\vec S}}_{-\vec Q}(0) \rangle
\langle {\delta {\hat N}}_{\vec Q}(0) {\delta {\hat N}}_{-\vec Q}(0) \rangle,
\end{equation}
including  purely spin correlation functions determining  both the uniform $(\vec Q =0)$ and staggered $(\vec Q \not= 0)$ purely spin susceptibility. In this connection, note an appearance of the anomalously large "induced" spin susceptibility as a result of the similar "antiferromagnetic" fluctuations both in purely spin and charge subsystems. Actually, this corresponds to an appearance of the hybrid {\bf ferrimagnetic fluctuations} (see Fig.10).
 Peculiar feature of the induced ferrimagnetic contribution  $\Delta \chi$ to the paramagnetic susceptibility is its anomalous behavior near the temperature $T_{CO}$ of the charge ordering in the local boson system, schematically pictured in Fig.9. Note the "pseudo-gap" behavior
at $T<T_{CO}$, revealing in a relatively sharp decrease of the  $\Delta \chi$ with decreasing temperature below $T_{CO}$. So, the temperature $T_{CO}$ of the charge ordering will determine an universal temperature behavior of the  induced ferrimagnetic contribution  $\Delta \chi$ to the paramagnetic susceptibility typical for the critical regions.

An interesting peculiarity of the  induced spin fluctuations in the "bi-plane" cuprates like
$YBa_{2}Cu_{3}O_{6+x}$ is an appearance of the  $q_{z}$-dependence of the correlation function (31) even in the absence of the purely spin exchange coupling between the nearest $CuO_{2}$-planes.
Really, the purely spin exchange interaction between the nearest
$CuO_{2}$-planes is relatively weak that implies the possibility to neglect the terms with   $\mu \not= \mu'$ in (31) and so, in practice, the spin correlation functions like $\langle {\hat {\vec S^{\mu}}} _{{\vec q}} {\hat {\vec S^{\mu'}}}_{-{\vec q}'} \rangle$ for the PJT centers system  do not depend on $q_{z}$.
On the other hand, the  long range inter-plane Coulomb coupling and the appropriate boson transfer are relatively large that results in an essential $Q_{z}$-dependence of the charge correlation functions like
$\langle {\delta {\hat N}}^{\nu} _{{\vec Q}} {\delta {\hat N}}^{\nu'} _{-{\vec Q}'} \rangle$.
Namely, this effect for the bi-plane cuprates is accompanied by a separation of the charge modes to the acoustical and optical ones in optimally doping compounds. So, these modes will  be revealed in the induced spin fluctuations determining the corresponding
$q_{z}$-dependence of the correlation function like
$\langle {\delta {\hat {\vec S}}_{\vec q}} (0) {\delta {\hat {\vec S}}_{-\vec q}} (t) \rangle$ $\sim \sin ^{2}{(\frac{q_{z}d}{2})}$ or  $\cos ^{2}{(\frac{q_{z}d}{2})}$ ($d$ is the distance between the $CuO_{2}$ planes).

For the description of the ${\vec q},  \omega$-dependencies of the induced spin fluctuations it may be used approximations like the known Ornstein-Zernike  model  applicable to the systems with well developed fluctuations.
Within a semiempirical approach we represent correlation function (29) in the integral form as
\begin{equation}
\langle {\delta {\hat {\vec S}}_{\vec q}} (0)	{\delta {\hat {\vec S}}_{-\vec q}} (t) \rangle
=
\int
P^{2}_{S}({\vec Q})
\langle {\hat {\vec S}}_{{\vec q}-{\vec Q}} (0) {\hat {\vec S}}_{-{\vec q}+{\vec Q}} (t)
{\delta \hat N}_{\vec Q} (0) {\delta \hat N}_{-\vec Q} (t)  \rangle d{\vec Q} ,
\end{equation}
where for the spin and charge correlation functions we can make use of the Lorentz or Gauss model form
$$
\langle {\hat {\vec S}}_{\vec q} {\hat {\vec S}}_{-\vec q} \rangle  \sim
\Biggl \{
\begin{array}{c}
\frac{1}{1+\xi^2_{sp}(\vec q - \vec q_{sp}^{(0)})^2}\\
e^{-a_{sp}(\vec q - \vec q_{sp}^{(0)})^2} \\
\end{array},
$$

$$
\langle {\hat \delta N}_{\vec Q}  {\hat \delta N}_{-\vec Q} \rangle \sim
\Biggl \{
\begin{array}{c}
\frac{1}{1+\xi^2_{ch}(\vec Q - \vec Q_{ch}^{(0)})^2}\\
e^{-a_{ch}(\vec Q - \vec Q_{ch}^{(0)})^2}\\
\end{array}.
$$

Here $\xi_{sp}$, $\xi_{ch}$ $(a^{-\frac{1}{2}}_{sp}$, $a^{-\frac{1}{2}}_{ch})$ are the correlation lengths for the spin and charge fluctuations, respectively.
 The resultant $q$-dependence of the
induced spin fluctuations can have a remarkable variety of forms exhibiting the features of both spin and charge fluctuations. Fig.11 shows some typical curves modelling in the Gauss approximation the well known "four-peak" structure of the dynamical spin susceptibilities revealed by the inelastic magnetic neutron scattering  for the cuprates like $La_{2-x}Sr_{x}CuO_{4}$  \cite{Pekin}. Incidentally, we conjectured an occurence of the antiferromagnetic singlet-triplet fluctuations with the specific wave vector ${\vec q}_{sp}=(1,\,1)$ and an appearance of the the substitution driven charge or charge-vibronic incommensurate modes with $q_{ch}=(\pm 2\epsilon , 0)$ or $(0, \pm 2\epsilon)$ compatible with the stripe like superstructures with $\epsilon \approx x$ \cite{Pekin}.

\section{The spin statics and dynamics in cuprates.}

An analysis of the various experimental data on magnetic susceptibility, neutron scattering, magnetic resonance (ESR,  $\mu$SR, NQR, NMR), Raman effect in the different cuprate HTSC's shows an absence of the long range magnetic order along with an existence of the well developed spin fluctuations and sufficiently strong exchange interactions. As a rule, both static and dynamical spin susceptibility probed by different methods exhibit more or less  distinctly a critical or "gap" behavior near the superconducting transition temperature $T_c$ and/or at a noticeably more higher temperature $T^{\ast}$ usually called as the "pseudo-gap" opening temperature. This phenomenon evidences the strong spin-charge intermode coupling. However, in some cases the superconducting transition doesn't result in either observable changes of the spin statics and/or dynamics.

A complicated unconventional behavior of the spin system in cuprates can be in natural way explained at least qualitatively within the developed singlet-triplet PJT centers model. Below, within a framework of this model we'll shortly discuss a number of the most significant and hotly debated experimental results.

{\bf Static spin susceptibility.} As an example we'll consider the temperature and concentration behavior of the  static spin susceptibility $\chi _S$ in $La_{2-x}Sr_{x}CuO_{4}$ \cite{Nakano,Nakano1} which is a simplest model system among the HTSC's. The Fig.12  schematically shows the typical $\chi _{S}(T)$ dependences for a number of the 214-systems based on the experimental data \cite{Nakano,Nakano1}  for $x$ in the range $0.06\div 0.26$. These can be reasonably explained in the framework of the model approach to the cuprates as to the singlet-triplet systems with strong fluctuations of the singlet-triplet separation $\Delta _{ST}$ at the positive sign and relatively small magnitude of the corresponding average value $\bar \Delta _{ST}\,>\,0$ or, in other words, at the predominantly singlet bare ground state of the PJT centers. Character of the temperature dependences $\chi _{S}(T)$ for different $x$ distinctly evidences the rise of these singlet-triplet fluctuations with the increase in concentration $x$ accompanied by the possible decrease of the appropriate $\bar \Delta _{ST}$ value. Moreover, this explains a decrease in the $T_c$ value at $x\,>\,x_{opt}\approx 0.15$ due to more and more effective spin reduction of the local boson transfer integral. In terms of the above developed singlet-triplet model  the unusual  temperature and concentration dependences of the paramagnetic susceptibility for $La_{2-x}Sr_xCuO_4$ in a wide range of temperatures $T>T_c$ and concentrations $0.06\div 0.26$  can be easily explained by the occurence of the $SS-VV$ crossover regime for the singlet-triplet fluctuations. Observed decrease of $\chi _{S}$ with lowering the temperature below $T_{max}(x)$ can be coupled with the opening of the spin gap (see the inset to Fig.5) in accordance with the conjecture of the authors  \cite{Nakano} that "the exceeding decrease of $\chi _{F}^{s}(T)$ at low temperatures will be due to a development of some kind of singlet state".
Note that an appearance of the appreciably large Curie term in the temperature behavior of the spin susceptibility $\chi _{S}$ (for example in $La_{2-x}Sr_xCuO_4$ above $x>0.2$ \cite{Nakano} or in $Tl_{2}Ba_{2}CuO_{6+x}$ \cite{Vyaselev}) can be linked not with paramagnetic impurities but with an appreciable weight of fluctuations resulting in the paramagnetic "$S$-type" contribution of the  PJT centers.

The minimal spin reduction of the local boson transfer integral essential for the optimized systems \cite{isotope} presumes a stability of the spin state of the PJT centers with regard to charge fluctuations. In this connection, one should pay attention to the  temperature stable systems with the temperature independent normal state spin susceptibility as to the possible optimized systems with maximal $T_c$. In such systems  change in the temperature or a heat fluctuation doesn't accompanied by an appreciable change of the local spin state. Some authors \cite{Triscone} consider condition $\frac{d\chi _S}{dT}\approx \,0$ as essential signature of the optimized systems like $YBa_{2}Cu_{3}O_{6.96}$.
Note the interesting analogy with the vibronic reduction of the local boson
 transfer integral and condition of the minimal isotope-effect
and baric coefficient $\frac{dT_c}{dp}$ for the optimized
systems \cite{isotope}.

{\bf The Knight shift.} The spin part of the Knight shift $^{n}K$ in the framework of the singlet-triplet model unlike the usually applied single-component spin model \cite{MMP} includes two contributuions connected with the usual spin susceptibility $\chi _S$ and unconventional  susceptibility $\chi _V$. So, for the copper nuclei $^{63,65}Cu$ the taking account only of the direct hyperfine coupling gives
\[
^{63,65}K_{sp}=\frac{1}{2\mu _{B}N}A_{Cu}(\chi _{S}+\chi _{V}),
\]
where $A_{Cu}$ is the direct hyperfine coupling parameter, $\mu _{B}$ is the Bohr magneton, and $N$ is the Avogadro's number. On the other hand, for the $^{17}O$ nuclei the taking account of the direct hyperfine coupling with the $e_u$-orbital gives
\[
^{17}K_{sp}=\frac{1}{2\mu _{B}N}A_{O}(\chi _{S}-\chi _{V}),
\]
where $A_{O}$ is the effective  hyperfine coupling parameter. So, a comparison of the temperature dependences of the $^{63,65}K$ and $^{17}K$ allows to examine the singlet-triplet model  and separate the $S$- and $V$-contributions. As an illustration note the observed discrepancy between the temperature dependences of the $^{63,65}K$ and $^{17}K$ in $La_{2-x}Sr_xCuO_4$ \cite{Nakano2} that can be considered as a striking evidence in favour of the two-spin-component singlet-triplet model with the $S$ and $V$ contributions to the Knight shift. Note, that to a certain extent taking account of $S$ and $V$ terms is equivalent to that of localized copper spin and delocalized oxygen hole spin contributions sometimes enlisted for detailed analysis of hyperfine coupling and nuclear resonance in cuprates \cite{Barrett}.

{\bf Spin relaxation.} Experimental studying the low-frequency spin dynamics in cuprates is mainly provided by the measurements of the NQR, NMR spin relaxation rates for different nuclei and the ESR spin relaxation rates for the rare earth ions (first of all for $Gd^{3+}$).

Spin  relaxation rates are mainly determined by the long range and long lived fluctuations that implies their critical (or near-critical) behavior near both the uniform phase transition temperatures like $T_N,\,T_{CO},\,T_{BS},\,T_c$ and percolative phase transitions. Especial attention should be paid  to the phase separation regime when a  phase transition within single phase is obligatory accompanied by the long range and long lived fluctuations of the coexisting phases.

A specific for the local boson model phase transition at $T=T_{CO}$ with the charge ordering at ${\vec Q}=(1,\,1)$ should be revealed in a critical behavior of the  spin-lattice relaxation rate $T_{1}^{-1}$ only for the copper nuclei $^{63,65}Cu$, but not for the nuclei like $^{17}O$, which crystallographic position results in a cancelation of the "antiferromagnetic" like fluctuations contribution. Indeed, an expected critical behavior of $T_{1}^{-1}$ (or $(T_{1}T)^{-1}$) for  the copper nuclei $^{63,65}Cu$ has been observed in $La_{2-x}Sr_xCuO_4$ ($T_{CO}\approx 50\,K$ for $x=0.13$ and $x=0.18$ \cite{Itoh}), in $YBa_{2}Cu_{3}O_{6.63}$ ($T_{CO}\approx 150\,K$ \cite{MMP}), in optimized $YBa_{2}Cu_{3}O_{6.96}$, where $T_{CO}\approx T_{BS}\approx T_c$ \cite{MMP}.
Appropriate temperature dependences for the $^{63}(T_{1}T)^{-1}$ within the above developed induced spin fluctuations model in accordance with expressions like (29) are determined by the temperature dependences of the charge fluctuations $\langle {\delta {\hat N}}_{\vec Q} {\delta {\hat N}}_{-\vec Q} \rangle _{\omega \rightarrow 0}$ and look like those shown in Fig.9.

A relatively low value of the  spin-lattice relaxation rates for the $^{17}O,\,^{89}Y,\,^{139}La$ and their temperature dependences can be qualitatively explained within a singlet-triplet model with assumptions used in the case of static spin susceptibility.

{\bf Inelastic neutron scattering.}
Magnetic inelastic neutron scattering is a powerful tool for
studying the detailed $\omega , {\vec q}$-dependences of
spin fluctuations. An appearance of new phase of the
PJT centers is accompanied by a dramatic modification in
magnetic excitations with large broadening in wave vector and
substantial redistribution of spectral weight in frequency \cite{Hayden}.
Magnetic fluctuations extend to a wide spectral range up to energies
$\sim 0.3\,eV $ with spectral weight removed from
the appropriate magnetic Bragg peak observed for the parent
antiferromagnetic compounds. To a certain extent the
magnetic response for the basic model HTSC's $La_{2-x}Sr_xCuO_4$
and $YBa_{2}Cu_{3}O_{6+x}$ can be represented
as a superposition of a strong broad background due
to antiferromagnetic fluctuations with ${\vec Q}=(1,\,1)$
and the relatively weak and sharp low frequency
generally incommensurate features peaked near $10\,meV$
in 214 system and near $30\div 40\,meV$ in 123 system.
A structure of the basic "antiferromagnetic" background
changes with concentration $x$ exhibiting a phase
separation regime with a suppression of the
parent phase contribution and a corresponding rise of the PJT centers phase singlet-triplet
contribution. Antiferromagnetic correlation length for
the former  decreases with doping from relatively large
values for slightly doped compounds to minimal values
$\xi \sim R_{CuCu}$, whereas that of the latter  remains minimal indicating
a permanently short-range character of the antiferromagnetic singlet-triplet fluctuations.
As seen from the experimental data  \cite{Hayden} a normalized
intensity of the magnetic inelastic neutron scattering for the
PJT centers phase is in common sufficiently weaker than that for the parent
antiferromagnetic phase at least at $T\,<\,300\,K$.
This is entirely compatible with the qualitative conclusions of the singlet-triplet
model with predominantlty singlet ground state of the PJT centers.
Unfortunately, the experimental data for the $YBa_{2}Cu_{3}O_{6+x}$
are limited to relatively low frequencies ($\hbar \omega \sim 120\,meV$) that doesn't permit to correctly examine the singlet-triplet contribution. Nevertheless, the available magnetic INS-data for various doping level  agree with a superposition model within phase separation regime and  can be effectively treated in framework of the {\it two Lorentz oscillators model} \cite{Pekin}.

{\bf Phase separation and magnetic properties of cuprates.} Phase separation both chemical and physical should be taken into account for a correct interpretation of the available experimental data. Properly speaking, within a phase separation regime we deal with contributions of two (or even more) phases with varying relative volume. In many cases, there are some difficulties with a distinct separation between single-phase and multi-phase effects. In particular, it concerns the phase transition phenomena accompanied by the varying the relative phase volume. Moreover,  a phase separation regime can exhibit a specific phase transition of the percolative origin.

Below, we list shortly some experimental results of the static and dynamic magnetic measurements which more or less distinctly indicate an occurence of the phase separation regime in either cuprates.

As one of the recent remarkable experimental indications to the nucleation  of the PJT centers phase in the cuprates note the zero field copper NMR data in $Y_{1-x}Ca_{x}Ba_{2}Cu_{3}O_6$ by P. Mendels et al. \cite{Mendels}. The nonisovalent substitution in the antiferromagnetic state was accompanied by the anomalous decrease in the concentration of the NMR resonating copper nuclei: every $Ca^{2+}$-ion excluded from the NMR about 50 copper ions (!), that could be connected with their disproportionation within the new phase "droplets". A decrease in the volume of the antiferromagnetic phase in $YBa_{2}Cu_{3}O_{6+x}$ with increasing $x$ results in a suppression of the Bragg peaks intensity in the inelastic neutron scattering spectra \cite{neutron}.

An occurence of two characteristic temperatures, namelly, that of the charge ordering ($T_{CO}$) and appearance of the local superconducting order ($T_{BS}$) gives rise for the two-dimensional regions of the PJT centers phase within the $CuO_2$-planes. It can be revealed in many effects even far from the percolative transition to the superconducting state $x<0.06$ for $La_{2-x}Sr_xCuO_4$ and $x<0.45$ for $YBa_{2}Cu_{3}O_{6+x}$ or within a formally antiferromagnetic dielectric state. So, a well known anomalous temperature behavior of the Bragg peak intensity and spin-echo relaxation rate in $YBa_{2}Cu_{3}O_{6+x}$ \cite{neutron,Matsumura} ("reentrant transition"), schematically shown in Fig.13, can be explained by a decrease in a volume of the antiferromagnetic regions with decreasing the temperature below $T\sim 30\,K$($T_{BS}$?). A similar effect is revealed by an anomalous  temperature dependence of the Zeeman splitting in the $^{139}La$ NQR spectra for $La_{2-x}Sr_xCuO_4$ at $T<\,30\,K$ \cite{Borsa}. A percolative nature of the superconducting transition in $La_{2-x}Sr_xCuO_4$ at $x\sim 0.06$ reveals in anomalous rise of the low temperature $^{139}La$ spin relaxation rates \cite{Kukovitsky}.

Interestingly, that strong isotop substitution effect in the spin susceptibility and the Neel temperature observed for $La_{2-x}Sr_xCuO_4$ \cite{Zhao} can be linked with an influence of the isotope substitution $^{16}O\rightarrow ^{18}O$ on the relative phase volume.

\section{Conclusions.}
A suggested singlet-triplet model considers the spin (magnetic) properties of the cuprates as an integral part of their multi-mode behavior characteristic for the systems with suficiently strong inter-mode coupling. Incidentally, it is rather simplified approach. In particular, we did not touch seriously upon the problem of the vibronic states and appropriate local structural Jahn-Teller modes. Actually, in cuprates we should deal with hybrid spin-charge-vibronic fluctuations.

Another important problem typical for the high-$T_c$ cuprates is connected with their quasi-two-dimensionality and its influence on the spin arrangement.

A supplementary important issue is coupled with the charge inhomogeneity and phase separation. These phenomena should be taken into account for an interpretation of the temperature and concentration dependencies of all both static and dynamical spin properties, especially, within the range of the critical behavior of either subsystem accompanied by the change of the relative weights and/or volumes of the coexisting phases.

Despite the used simplifications the singlet-triplet model enables to make some qualitative and semi-quantitative conclusions compatible with the observed experimental data. First of all, it concerns an occurence of the singlet-triplet antiferromagnetic fluctuations of various $S$ and $V$ types resulting in  various temperature dependence of different quantities describing the spin static and dynamic properties.  Let note, that unlike the widely popularized  nearly antiferromagnetic Fermi-liquid model \cite{MMP}
the suggested approach naturally enables to take account both of $d$- and $p$-holes in conditions of their comparable role.

As a whole,  singlet-triplet model can provide a promising foundation for the qualitative and semi-quantitative analysis of the static and dynamical spin properties of cuprates including paramagnetic susceptibility, nuclear resonance and  inelastic magnetic neutron scattering. Moreover,  the further elaboration of the singlet-triplet model represents a perspective task not only for the HTSC-problem.

\newpage{}

\begin{center}
	{\bf Figure captions.}
\end{center}

Fig.1. The structure of the lower energy levels for the PJT center. The four types of distorsions of the $CuO_4$ clusters corresponding to the four minima of the ground state adiabatic potential are schematically depicted in the insets for the $B_{1g}$ and $B_{2g}$ symmetry types, respectively.

Fig.2. The electron density distribution for the $^{1,3}E_{u}$ term of the
$b_{1g}e_{u}(\sigma)$  configuration (left) and the $b_{1g}e_{u}(\pi)$ one (right) compatible with the $B_{1g}$ type distorsion  of the $CuO_4$ clusters.
The dark filling corresponds to the $e_{u}$ orbitals and light filling does to  the $b_{1g}$ ones. The appropriate vibronic states  are ascribed to the $\sigma_{z}=\pm \frac{1}{2}$ pseudo-spin states.

Fig.3. Illustration to possible angular (a) and collinear (b,c) spin $S,V$ configurations for the singlet-triplet magnet.

Fig.4. The orientations of the copper and oxygen spins for the different  collinear phases (qualitatively). The distinguishing tone filling corresponds to the possible different orbital configurations.

Fig.5. The temperature dependence of the transversal (solid curve) and longitudinal (dotted curve)  spin susceptibilities $\chi _{S,V}$ and order parameters for the $SS$ phase ($\Delta _{ST} = 0.1 eV$, $z I^{(1)} = 0.4 eV$, $z I^{(2)} = - 0.3 eV$, $z I^{(4)} = 0.2 eV$). The inset shows an appropriate  simplified spin level scheme ($T=0$).

Fig.6. The temperature dependence of the transversal (solid curve) and longitudinal (dotted curve)  spin susceptibilities $\chi _{S,V}$ and order parameters for the $VV$ phase ($\Delta _{ST} = 0.1 eV$, $z I^{(1)} = 0.2 eV$, $z I^{(2)} = - 0.3 eV$, $z I^{(4)} = 0.4 eV$). The inset shows an appropriate  simplified spin level scheme ($T=0$) with the spin gap.

Fig. 7. (a) The temperature dependence of order parameters for noncollinear phase ($\Delta = 0.1 eV$, $z I^{(1)} = 0.2 eV$, $z I^{(2)} = 0.3 eV$, $z I^{(4)} = -0.1 eV$); (b) The temperature dependence of "intra-site" $\Theta_{AA} (T) = \Theta_{BB} (T)$ and "inter-site" $\Theta_{AB} (T)$ angles (see text).

Fig. 8. The temperature dependence of longitudinal (solid curve) and transversal (dotted curve) magnet susceptibilities $\chi _{s,v}$ for noncollinear phase ($\Delta = 0.1 eV$, $z I^{(1)} = 0.2 eV$, $z I^{(2)} = 0.3 eV$, $z I^{(4)} = -0.1 eV$).

Fig.9. (a): A model $(T,  \nu)$-phase diagram for the quantum lattice bose-gas ($N_{B} \sim  0.5$) with $\nu$ being a parameter (in particular $N_B$ or doping content!) determining a certain change of the boson-boson repulsion $V$ and the boson transfer integral $t$ (see the inset); (b,  c,  d) show a qualitative $T$-dependence of the charge fluctuations and the superconducting order parameter fluctuations for three regimes marked by the arrows in (a): $V > t$ (weak screening),  $V \approx t$ (optimal screening) and $V < t$ (over-screening). The standard abbreviations denote the non-ordered or metallic phase (NO), the charge order phase (CO), the bose-superfluid (superconducting) phase (BS), and mixed superconducting) phase (BS+CO).

Fig.10. An illustration to the induced ferrimagnetic ordering within the singlet-triplet model. Different filling for the $CuO_4$ centers indicates different charge (boson) density. For the sake of the close approach to the real cuprates like $La_{2-x}Sr_{x}CuO_{4}$ we indicate also the buckling of the $CuO_2$ planes (by different filling of the appropriate oxygen ions) and possible different sizes of the $CuO_4$ clusters with different charge  density.

Fig.11. The model "four-peak" structure of the dynamical spin susceptibility  for the cuprates like $La_{2-x}Sr_{x}CuO_{4}$ determined by induced spin fluctuations and revealed by the inelastic magnetic neutron scattering. Incidentally, it was  conjectured an occurrence of the antiferromagnetic singlet-triplet fluctuations with the specific wave vector ${\vec q}_{sp}=(1,\,1)$ and an appearance of the the substitution driven charge or charge-vibronic incommensurate modes with ${\vec q}_{ch}=(\pm 2\epsilon , 0)$ or $(0, \pm 2\epsilon)$ compatible with the stripe like superstructures with $\epsilon \approx  x$.

Fig.12. The typical $\chi _{S}(T)$ dependences for a number of the 214-systems based on the experimental data \cite{Nakano,Nakano1}. The curves 1-6 correspond to  $x$ increasing in the range $0.06\div 0.26$.
Arrows indicate $T_{max}$.
The inset shows a qualitative energy spectrum for the PJT centers with taking account of  the well developed fluctuations.

Fig.13.  Anomalous temperature behavior of the Bragg peak intensity and spin-echo relaxation rate in $YBa_{2-x}Cu_{3}O_{6.3}$) \cite{neutron,Matsumura} near $T=30\, K$ (schematically).
}

\begin{thebibliography}{99}

\bibitem{Moskvin} A.S. Moskvin,  JETP Lett.  58 (1993) 342.


\bibitem{Moskvin1}
A.S. Moskvin, N.N. Loshkareva, Yu.P. Sukhorukov, M.A. Sidorov, A.A. Samokhvalov,
JETP  105 (1994) 967.


 \bibitem{Moskvin2}
 A.S. Moskvin, Physica C (to be published).

\bibitem{Ber}
 I.B. Bersuker and V.Z. Polinger, Vibronic Interactions in Molecules and Crystals (Springer-Verlag, Berlin, 1989).

\bibitem{Fisk}  S. Yoshinari, P.C. Hammel, Z. Fisk, Phys. Rev. Lett. 77 (1996) 2016.

\bibitem{Kazan}
T.A. Ivanova, E.F. Kukovitsky, E.N. Nabiullin et al.,  JETP Lett.  57 (1993) 64.


\bibitem{robasz}
 S. Robaszkiewicz, R. Micnas, K.A. Chao,  Phys. Rev. B  23 (1981) 1447.


\bibitem{Alexandrov1}  A.S. Alexandrov, J. Ranninger, S. Robaszkiewicz,  Phys. Rev. Lett. 56 (1986) 949.


\bibitem{Bourges}
P. Bourges, H. Casalta, A.S. Ivanov and D. Petitgrand, Cond-mat/9708060 7 Aug. 1997.

\bibitem{birgeneau}  R. Birgeneau,  J. Als-Nielsen, E. Bucher,  Phys. Rev. B  6 (1972) 2724.

\bibitem{Hsieh}  Y.Y. Hsieh, M. Blume,  Phys. Rev. B  6 (1972) 2684.

\bibitem{oles}   A.M. Oles, L.F. Feiner, J. Zaanen,  Abstracts of ICM-94, Warsaw 22-26 August 1994, p. 257.


\bibitem{Bohm}
 A. Bohm,  Quantum Mechanics: Foundations and Applications (Springer, Berlin, 1986).

\bibitem{isotope}
 A.S. Moskvin, Yu.D. Panov, JETP 84 (1997) 354.


\bibitem{Pekin}  A.S. Moskvin,  A.S. Ovchinnikov, O.S. Kovalev,  Physica C (to be published).

\bibitem{Nakano}
T. Nakano, K. Yamaya, N. Momono, M. Oda and M. Ido, J. Low Temp. Phys.  105 (1996) 395.

\bibitem{Nakano1}
T. Nakano, N. Momono, C. Manabe, Y. Miura, M. Oda and M. Ido, Czech. J. Phys.  46, Suppl.2(1996) 1153.

\bibitem{Vyaselev}
O.M. Vyaselev, and I. Schegolev, Czech. J. Phys. 46, Suppl.2 (1996) 1137.

\bibitem{Triscone}
G. Triscone, J.-Y. Genoud, T. Graf et al., J. Alloys and Compounds, 195 (1993) 351.

\bibitem{MMP} A.J. Millis, H. Monien, D. Pines,
Phys. Rev. B 42 ( 1990) 167.

\bibitem{Nakano2}
T. Nakano et al.,  Phys. Rev. B  49 (1994) 16000.

\bibitem{Barrett}
S.E. Barrett, D.J. Durand, C.H. Pennington et al.,  Phys. Rev.B  41 (1990) 6283.

\bibitem{Itoh}
Y. Itoh, M. Matsumura,H. Yamagata, J. Low Temp. Phys. 105 (1996) 365.

\bibitem{Hayden}
S.M. Hayden, G. Aeppli, H.A. Mook et al., Phys. Rev. Lett. 76 (1996) 1344.

\bibitem{Mendels}
 P. Mendels, H. Alloul, G. Collins, G.F. Marucco, Physica C (to be published).

\bibitem{neutron}
J.M. Tranquada, G. Shirane, B. Keimer, S. Shamoto, M. Sato, Phys. Rev.B 40 (1989) 4503.

\bibitem{Matsumura}
M. Matsumura, H. Yamagata, Y. Yamada et al., J. Phys. Soc. Jap.  58 (1989) 805.

\bibitem{Borsa}
F. Borsa, P. Caretta, J.H. Cho et al., Phys. Rev.B  52 (1995) 7334.

\bibitem{Kukovitsky}
E. Kukovitsky, H. Lutgemeier, G. Teitel'baum, Physica C 252 (1995) 160.

\bibitem{Zhao}
Guo-meng Zhao, K.K. Singh, and D.E. Morris, Phys. Rev. B  50, 4112 (1994).

\end{thebibliography}
\end{document}